\documentclass[aps,twocolumn,pre,showpacs,eqsecnum]{revtex4-1}
\usepackage{graphpap}
\usepackage{graphicx}
\usepackage{graphics}
\usepackage{color}

\begin{document}

\title{Reducing disorder in graphene nanoribbons by chemical edge modification}
 \author{J. Dauber$^{1,2}$, B. Terr\'es$^{1,2}$, C. Volk$^{1,2}$, S. Trellenkamp$^2$, and C. Stampfer$^{1,2}$}
 \affiliation{
$^1$\,JARA-FIT and II. Institute of Physics, RWTH Aachen University, 52074 Aachen, Germany, EU\\
$^2$\,Peter Gr\"unberg Institute (PGI-8/9), Forschungszentrum J\"ulich, 52425 J\"ulich, Germany, EU
}

\date{ \today}

\begin{abstract}
We present electronic transport measurements on etched graphene nanoribbons on silicon dioxide 
before and after a short hydrofluoric acid (HF) treatment. We report on changes in the
transport properties, in particular, in terms of a decreasing transport gap and a reduced 
doping level after HF dipping. Interestingly, the effective energy gap is nearly unaffected by the HF treatment.
Additional measurements on a graphene nanoribbon with lateral graphene gates support strong indications that the HF 
significantly modifies the edges of the investigated nanoribbons leading to a significantly reduced
disorder potential in these graphene nanostructures.

\end{abstract}

 \pacs{???}  
 \maketitle
\newpage

A major challenge  for implementing
 semiconductor device-concepts in graphene is the absence of an energy gap. This problem can be circumvented by etching nanostructures in bulk graphene, 
which introduces a size-dependent effective energy gap, 
as observed in a number of experiments on 
graphene nanoribbons (GNRs)~\cite{che07,han07,dai08,wan08,mol09,sta09,tod09,liu09,mol10,gal10,han10,ter11,mor10}.
The transport properties of these nanoribbons appear however to be 
  heavily influenced by disorder and localized states ~\cite{sta09,tod09,liu09,mol10,gal10}. 
The main contributions to disorder are 
 most likely due to: (i) interaction with the substrate, 
 (ii) edge roughness (including dangling bonds or chemical edge modifications) and (iii) impurities on graphene. The latter problem can be addressed by annealing techniques. Substrate 
induced disorder can be strongly reduced by placing 
graphene on hexagonal boron nitride (hBN). However, while this significantly improves the properties of bulk graphene~\cite{dea10,xue11,for13},  graphene nanostructures on hBN have shown very similar transport characteristics compared to those on silicon dioxide (SiO$_2$)~\cite{bis12,eng13}. 
Very similar findings have also been made
for suspended graphene nanoribbons~\cite{tom11,mor12}. 
These measurements provide strong indications, 
that edge roughness 
is the dominant contribution to disorder in 
graphene nanostructures, limiting their transport properties. These observations are also consistent with the
improved transport properties of nanoribbons made from unzipping carbon nanotubes~\cite{dai08,wan11}, where cleaner
edges are expected. Edge modifications of etched graphene nanostructures 
 might therefore be a very important route for improving the transport properties of these devices.

Here, we present electronic transport measurements on etched GNRs before and after a short treatment with lowly concentrated hydrofluoric acid (HF). A short etching time is chosen to have only chemical interaction of the HF with the edge of the GNRs without actually suspending the structure. We 
investigate several devices and, for each of them, compare the transport properties before and after the HF treatment.
We show that the transport gap as well as the doping level are significantly reduced 
by the HF dip, while the effective energy gap is roughly unaffected. 
Moreover, we consider the effects of side gates on a nanoribbon that underwent a HF dip. By comparing the results from similar samples, which have not been exposed to HF, we gain strong indications that edge modification plays a major role in the HF treatment. 

\begin{figure}[tb]\centering 
\includegraphics[draft=false,keepaspectratio=true,clip,%
                   width=0.93\linewidth]%
                   {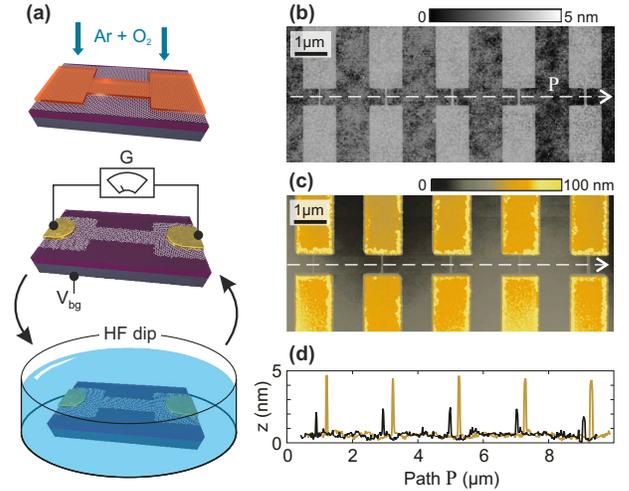}                   
\caption{
(a) Illustrations of fabrication and measurement steps: Patterning the GNRs by Ar/O$_2$ plasma, conductance measurements and HF dipping. (b) SFM image of an array of etched GNRs before contacting. The length of the GNRs are 500~nm and the width vary between 40 and 80~nm. (c) A SFM image of the same structure after metal contacting and HF dipping. (d) Cross sections of the SFM images above, (see dashed lines in panels (b) and (c)), showing the GNRs height before (black) and after (brown) HF treatment. A horizontal offset of 300~nm is added for clarity.
}
\label{processflow}
\end{figure}


The graphene samples have been fabricated by mechanical exfoliation of natural graphite. Graphene flakes are deposited onto 285~nm thick SiO$_2$ on a highly p-doped Si substrate and their single-layer nature is confirmed by Raman spectroscopy~\cite{dav07a}. For structuring GNRs, an etch mask of polymethyl methacrylate (PMMA) resist is patterned by electron beam lithography (EBL) and uncovered graphene is removed by reactive ion etching based on an Ar/O$_2$ plasma (see illustrations in the top-panels of Fig.~1(a)). A scanning force microscope (SFM) image of several etched GNRs is shown in Fig.~1(b). 
Contacts have been made by a second EBL step based on PMMA resist, metal evaporation (5~nm Cr/50~nm Au) and a standard lift-off.
After a first electrical characterization of the individual GNRs, the samples are dipped in 1\% HF solution for 20~s and cleaned with deionized water. Fig.~1(c) shows a SFM image of the same GNRs as in Fig.~1(b) after contacting and HF dipping. 
The HF etching approximately removes 3-5~nm of SiO$_2$, which is  
not leading to suspended nanoribbons.
A cross-section of the data of Figs.~1(b)-(c) along the indicated dashed lines is shown in Fig.~1(d). 
The observed step height difference of about 2-3~nm is consistent with the expected etching rate.

\begin{figure}[t]\centering
\includegraphics[draft=false,keepaspectratio=true,clip,%
                   width=\linewidth]%
                   {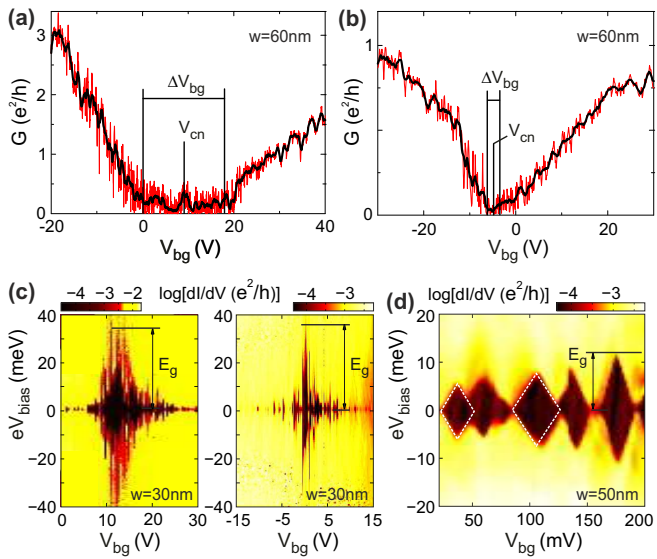}                   
\caption[FIG1]{
(a) and (b) Back gate characteristics at $V_{\rm bias}=300~\mu$V of a GNR of 
width $w=$~nm before (a) and after (b) HF dipping. The red curve shows the raw data and the black curve represents a running average over 0.5~V. (c) Color plot of differential conductance $dI/dV$ as function of source-drain bias and back gate voltage before (left) and after (right) HF dipping. These data correspond to a nanoribbon of width $w=30$~nm. 
(d) High resolution Coulomb diamond plot of a GNR of width $w=50~$nm. Dashed lines represent an estimate of the diamonds for the lever  arm analysis. All devices considered in this figure have the same length $l=500$~nm.
}
\label{gaps}
\end{figure}

To investigate the effects of the HF treatment, we perform transport measurements before and after HF dipping on the same devices.  All transport measurements are carried out 
in a pumped $^4$He system at $T\approx$~1.4~K in He atmosphere. The current through a nanoribbon is measured 
by applying a dc bias voltage symmetrically to source and drain. 
The differential conductance is measured directly using low-frequency lock-in techniques by adding a small ac bias of 100~$\mu$V. The highly doped silicon substrate is used as a global back gate (BG) for tuning the carrier density in the nanoribbons. 

There are two relevant quantities 
that characterize the transport through a GNR. One 
is the so-called {\em transport gap} $\Delta V_{\rm bg}$, which is the region of back gate voltage 
where the low-bias conductance is strongly suppressed.
The other  is the {\em effective energy gap} $ E_{\rm g}$, which is the maximum range of suppressed conductance in bias voltage direction. Finally, another important observable is the charge neutrality point $V_{\rm cn}$, which is defined as the center of the transport gap, and is related to the doping level of the sample. 

These three quantities are affected differently by the HF treatment, as illustrated in Fig.~2. The transport gap $\Delta V_{\rm bg}$ is strongly suppressed  after the HF dipping, see Figs. 2(a) and 2(b). Here we show low-bias conductance as function of back gate voltage $V_{\rm bg}$ recorded on a nanoribbon with a width of $60$~nm and a length of $500$~nm before (Fig.~2(a)) and after (Fig.~2(b)) HF dipping. Both measurements exhibit an ambipolar transport characteristic, where 
the transport gap $\Delta V_{\rm bg}$, separates the hole- (left) from the electron-dominated (right) transport region. 
Quantitatively estimating $\Delta V_{\rm bg}$ ~\cite{com02}, we obtain $\Delta V_{\rm bg} \approx 18.3$~V  
before and $\Delta V_{\rm bg} \approx 2.8$~V after HF treatment, \textit{i.e.} the transport gap is reduced roughly by a factor of 6. Together with this suppression, 
the charge neutrality point $V_{\rm cn}$  
is shifted closer to zero BG voltage, indicating a reduction of doping after HF dipping. In addition, the overall conductance level is changed, which we mainly attribute to an increase of the contact resistance due to oxidation of the chromium adhesion layer in ambient conditions during the additional processing.

\begin{figure*}[t]\centering
\includegraphics[draft=false,keepaspectratio=true,clip,width=0.96\linewidth]{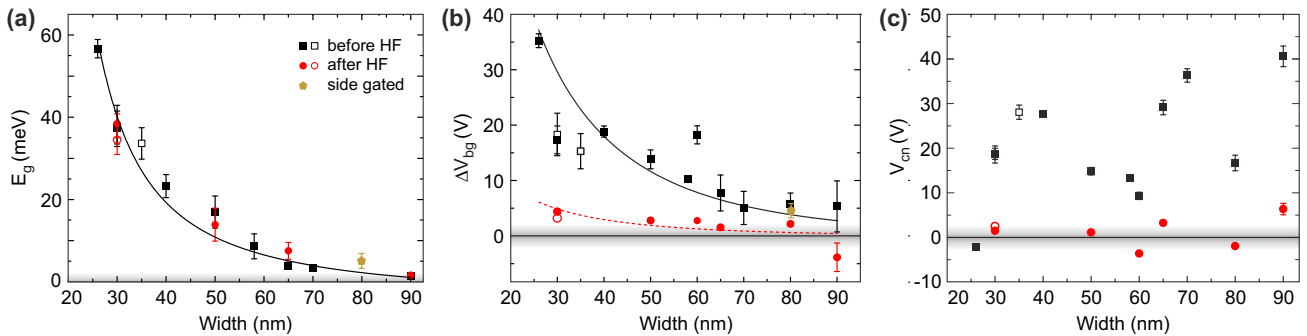}
\caption[FIG1]{Results from the measurements of several GNRs [length $l=500$~nm (filled symbols) and $l=200$~nm (open symbols)], before and after HF dip. (a) Effective energy gap $ E_{\rm g}$ before and after HF dip.  The data are fitted with the two-parameter model  of Sols \textit{et al.}~\cite{sol07}, $ E_{\rm g}(w)=a/w \cdot \exp(-b w)$. For our data we find $a=2.2$~eV~nm and $b=0.021$~nm$^{-1}$,  in good agreement with earlier observations~\cite{comCS01}.
(b) Transport gap $\Delta V_{\rm bg}$ before and after HF dip. In agreement with Ref.~\cite{mol10},  $\Delta V_{\rm bg}$ is approximately proportional to $E_{\rm g}$, \textit{i.e}.  $ \Delta V_{\rm bg}(w)=\beta_{b(a)}\cdot E_{\rm g}(w)$, with $\beta_{b} \approx0.64$~V/meV  
before the dip (black line), and $\beta_{a}=\beta_{b}/6$ after the dip (red dashed-line). Interestingly,  $\beta_{b}$ coincides with the value of Ref.~\cite{mol10}.
(c) Position of the charge neutrality point $V_{\rm cn}$ with respect $V_{\rm bg}$=0~V before and after HF dip.
}
\label{results}
\end{figure*}

The effective energy gap $E_{\rm g}$ of a nanoribbon can be extracted by measuring the differential conductance ($dI/dV$) as function of source-drain and BG voltage, as it  corresponds to the maximum span in bias direction of the region of strongly suppressed conductance, 
see Fig.~2(c). 
Interestingly, the value of the effective energy gap $ E_{\rm g}$ is approximately the same before and after HF dipping. 
This observation contrasts to the behavior of the transport gap and the doping level which, also for this device (width 30~nm, length 500~nm), shows a significant suppression after HF treatment. 

To understand the effects of HF treatment, another important quantity to consider is the back gate lever arm $\alpha$, which is a measure for the capacitive coupling of the BG to  individual charge islands in the nanoribbon~\cite{sta09}. This can be extracted by analyzing the slope of the edges of the Coulomb 
diamonds that characterize the regime of suppressed conductance~\cite{mol10}. These diamonds are clearly distinguishable in high-resolution close-ups  of the  differential conductance $dI/dV$, see Fig.~2(d).  
A systematic analysis of similar measurements on a number of different nanoribbons  before and after HF dipping gives an averaged lever arm  $\alpha_b \approx 0.21$ before and $\alpha_a \approx 0.24$ after HF treatment \cite{footnote1}. These values are consistent with earlier measurements on similar structures~\cite{mol10,gal10,mor10} and clearly indicate that the electrostatic coupling to the back gate is not affected 
significantly by the HF treatment. This allows to associate the observed reduction of the transport gap $\Delta V_{\rm bg}$ to a suppression of the disorder potential in the nanoribbon after HF dipping.

In Fig.~3 we summarize measurements similar to those
shown in Fig.~2 for a number of nanoribbons with different
widths.
Again, we observe that the effective energy gap is almost unchanged after the HF treatment, while the transport gap is significantly suppressed and the charge neutrality point is shifted towards zero back gate voltage. Note that also the scattering of the data points in Figs.~3(b)-(c) is significantly reduced for samples that underwent HF dipping. All these observations point to the conclusion that the HF treatment promote a significant suppression of the disorder potential in the nanoribbon.  In fact, earlier observations~\cite{bis12} suggests that the transport gap $\Delta V_{\rm bg}$ is sensitive to microscopic details of the edge disorder, which is in turn related to doping and to the presence of localized states.  Vice versa, the effective energy gap $E_{\rm g}$ is expected to be only weakly affected by disorder, as it is related to the charging energy of the smallest charge island along the GNR~\cite{sta09,tod09} and it is primarily set by the width of the nanoribbon itself~\cite{sol07}.

In principle, there are (at least) three possible ways in which HF dipping can affect the disorder potential of graphene nanoribbons: (i) it  may remove impurities from the GNRs and their surroundings and by that, lowering the doping and the disorder level; (ii) it might reduce the surface-induced disorder by modifying the graphene-SiO$_2$ interface; and (iii) it might alter the edges of the nanoribbons.  
However, it is known from suspended graphene devices that such HF cleaning is not very efficient, since advanced annealing techniques are required to obtain low doping and high charge carrier mobilities~\cite{du08,bol08}. Improvements of the graphene-substrate interface are also not expected to account for the significant reduction of the transport gap observed in our experiments, as the conduction 
properties of narrow GNRs have shown little enhancement even on high-quality hBN~\cite{bis12,eng13}. This makes a modification of the edges of the graphene nanoribbons a primary candidate to explain 
 the effects we observe.  

\begin{figure*}[t]\centering
\includegraphics[draft=false,keepaspectratio=true,clip,%
                   width=1.0\linewidth]%
                   {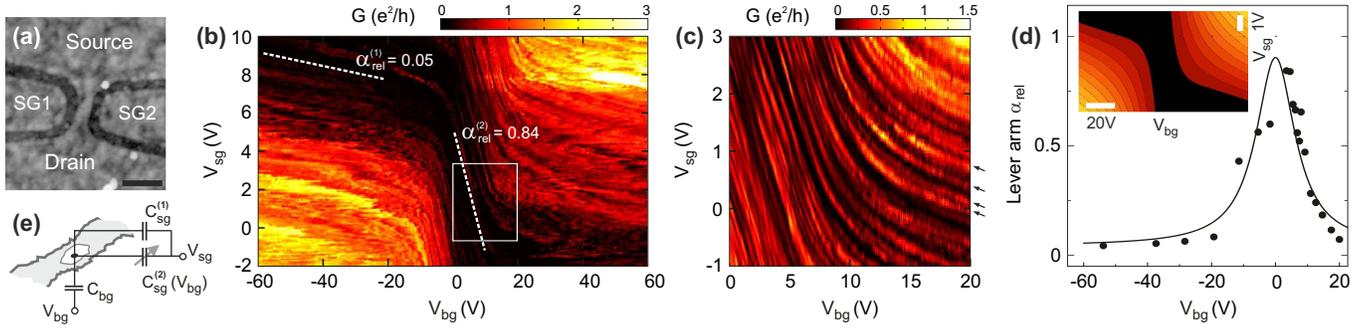}                   
\caption[FIG1]{
(a) SFM image of a graphene nanoribbon with side gates (scale bar of 200 nm). The GNR is about 80~nm wide and 120~nm long. The graphene side gates (SG1 and SG2) are approximately 40~nm separated from the nanoribbon. 
(b) Color-scale plot of the conductance of the nanoribbon after HF dipping as function of back gate and side gate voltage. White dashed lines indicate the relative lever arm in two different regions. (c) Close-up of the region highlighted by the white box in panel (b). 
(d) Relative lever arm $\alpha_{\rm rel}$ as a function of BG voltage for $V_{\rm sg}$=0~V. Solid line is based on a model of two parallel capacitances (see text).  Inset: Modeled conductivity of $\sigma(V_{\rm bg},V_{\rm sg}) = e\cdot n(V_{\rm bg},V_{\rm sg})\cdot\mu$ in arbitrary units (constant mobility $\mu$). The charge carrier density has been modeled as $n(V_{\rm bg},V_{\rm sg})=\alpha_0 [V_{\rm bg} + \alpha_{\rm rel}(V_{\rm bg}) \cdot V_{\rm sg}]$, with constant $\alpha_0$.  
(e) Schematic illustration of the capacitance model.
}
\label{S-shape}
\end{figure*}
 
Further indications of the influence of the HF treatment on the edges of the nanoribbons come from measurements on a GNR with lateral side-gates shown in Fig.~4(a). 
In Fig.~4(b) the conductance of a device, that was exposed to HF treatment, is shown as a function of the voltage applied to the back gate, $V_{\rm bg}$ and to the side gates $V_{\rm sg}=V_{\rm sg1}=V_{\rm sg2}$. 
The data show the cross-over from hole (lower left hand corner) to electron-dominated transport (upper right hand corner), separated by 
the transport gap of strongly suppressed conductance (dark area). In this region distinct resonance lines are clearly distinguishable, which are due to individual islands in the nanoribbon (see arrows in Fig.~4(c)). 
The slope of these features is a measure of the relative lever-arm of the back gate and side gates acting on the islands, $\alpha_{\rm rel}=C_{\rm bg}/C_{\rm sg}$.
In striking contrast to what has been observed in similar measurements on non-HF-treated GNRs (see \textit{e.g.} Fig.~4(b) in Ref.~\cite{mol09}), 
our experiment shows $\alpha_{\rm rel}$ is non-constant over the considered voltage range. Around $V_{\rm bg}=0$~V, we estimate $\alpha_{\rm rel} \approx$~0.84 (which is comparable with the value of 0.52 found
in Ref.~\cite{mol09}), while for  more positive or negative $V_{\rm bg}$ the relative lever arm is reduced to $\alpha_{\rm rel} \approx$~0.05. This behavior indicates that the capacitive coupling of the side gates to the islands in the nanoribbon has a contribution that depends on the BG voltage, 
$C_{\rm sg}(V_{\rm bg}) =C^{(1)}_{\rm sg} + C^{(2)}_{\rm sg}(V_{\rm bg})$, see Fig.~4(e). 
In Fig.~4(d) we show the relative lever arm $\alpha_{\rm rel}$ as function of BG voltage for $V_{\rm sg}=0$~V, and we consider $C^{(2)}_{\rm sg}(V_{\rm bg})= C_{0}[1-1/(1+\gamma V_{\rm bg}^2)]$ to describe our data points. Noticeably, using this simple model and the values for $C_{0}$ and $\gamma$ extracted from estimating $\alpha_{\rm rel}$~\cite{comm03} at $V_{\rm sg}=0$~V, we are able to qualitatively describe the behavior of the conductance in the whole voltage range, as shown by the inset in Fig.~4(d).

The different behavior of the sample of Fig.~4 with respect to nanoribbons that were not HF-treated~\cite{mol09}, 
is most likely due to edge modifications caused by the HF dipping. 
In fact, the density of states (DOS) in the near vicinity of the Fermi level is expected to get significantly altered, when hydroxyl (-OH) and oxygen (-O) terminated graphene is, at least partially, substituted by fluorine (-F) terminated graphene, which is likely to be favoured by the HF treatment~\cite{tac10,ihna11,zhe08}.   
By first-principle calculations it has been shown that replacing -OH and -O by -F terminations leads to a significant reduction of states
near the Fermi level. This reduction is mainly due to an extra electron provided by fluorine, which saturates the dangling carbon bonds along a zigzag edge~\cite{zhe08}.
Such a reduced DOS near the Fermi level may indeed, (i) cause an overall reduced disorder potential as discussed in Fig.~3 and
(ii) reduce the charge accumulation near the graphene edges. This results in tunable screening properties, allowing for a gate-controllable 
side-gate coupling.

Please be aware that a microscopic understanding of the edge modification and the origin of such a tunable side-gate coupling $C_{\rm sg}(V_{\rm bg})$ is still missing and requires further investigations. 
This may also be important for obtaining a higher yield in the presented process (see Fig.~3(b)).

In summary, we investigate etched graphene nanoribbons before and after a short HF treatment. We observe changes in the transport properties in terms of a decreased transport gap and a reduced doping level, but a nearly unaffected effective energy gap. The observations can be interpreted as a reduction of the disorder potential, whereas the width of the nanoribbon is still dominating the effective energy gap. Measurements on a side gated nanoribbon support indications that nanoribbon edge modifications play a very crucial role.
Although the mechanism behind this effect is related to the HF solution further details remain unknown. Advanced local experimental techniques 
and improved chemical routes are required for characterizing and improving the understanding of the graphene edge properties. 

{Acknowledgments ---}
We thank A.~Steffen, R.~Lehmann, H.-W.~Wingens, and U. Wichmann for help on sample fabrication and electronics,  B.~Beschoten and S.~Rotkin for discussions and A.~M\"uller and F.~Haupt for helpful input on the manuscript. Support by the HNF, JARA Seed Fund, the DFG (SPP-1459 and FOR-912) and the ERC are gratefully acknowledged.


\begin{thebibliography}{99}

\bibitem{che07}
Z. Chen, Y.-M. Lin, M. Rooks and P. Avouris, Physica E,~{\bf 40}, 228 (2007).

\bibitem{han07}
M. Y. Han, B. \"Ozyilmaz, Y. Zhang, and P. Kim, Phys. Rev. Lett.,~{\bf 98}, 206805 (2007).

\bibitem{dai08}
X. Li, X. Wang, L. Zhang, S. Lee, H. Dai, Science,~{\bf 319}, 1229 (2008).

\bibitem{wan08}
X. Wang, Y. Ouyang, X. Li, H. Wang, J. Guo, and H. Dai, Phys. Rev. Lett.,~{\bf 100}, 206803 (2008).

\bibitem{mol09}
F. Molitor, A. Jacobsen, C. Stampfer, J. G\"uttinger, T. Ihn, and K. Ensslin, Phys. Rev. B,~{\bf 79}, 075426 (2009).

\bibitem{sta09}
C. Stampfer, J. G\"uttinger, S. Hellm\"uller, F. Molitor, K. Ensslin, and T. Ihn, Phys. Rev. Lett.,~{\bf 102}, 056403 (2009).

\bibitem{tod09}
K. Todd, H.-T. Chou, S. Amasha, and D. Goldhaber-Gordon, Nano Lett.,~{\bf 9}, 416 (2009).

\bibitem{liu09}
X. L. Liu, J. B. Oostinga, A. F. Morpurgo, and L. M. K. Vandersypen, Phys. Rev. B,~{\bf 80}, 121407 (2009).

\bibitem{mol10}
F. Molitor, C. Stampfer, J. G\"uttinger, A. Jacobsen, T. Ihn, and K. Ensslin, Semicond. Sci. Technol.,~{\bf 25}, 034002 (2010).

\bibitem{gal10}
P. Gallagher, K. Todd, and D. Goldhaber-Gordon, Phys. Rev. B,~{\bf 81}, 115409 (2010).

\bibitem{han10}
M. Y. Han, J. C. Brant, and P. Kim, Phys. Rev. Lett.,~{\bf 104}, 056801 (2010).

\bibitem{ter11}
B. Terr\'es, J. Dauber, C. Volk, S. Trellenkamp, U. Wichmann, and C. Stampfer, 
Appl. Phys. Lett.,~{\bf 98}, 032109 (2011).

\bibitem{mor10}
J. B. Oostinga,  B. Sac\'ep\'e,  M. F. Craciun, and A. F. Morpurgo, Phys. Rev. B,~{\bf 81}, 193408 (2010).


\bibitem{dea10}
C. R. Dean, A. F. Young, I. Meric, C. Lee, L. Wang, S. Sorgenfrei, K.
Watanabe, T. Taniguchi, P. Kim, K. L. Shepard, and J. Hone, Nat. Nanotechnol.,~{\bf 5}, 722 (2010).

\bibitem{xue11}
J. Xue,	J. Sanchez-Yamagishi,	D. Bulmash, P. Jacquod,	A. Deshpande,	K. Watanabe,	T. Taniguchi,	P. Jarillo-Herrero	and B. J. LeRoy,
Nat. Matarials,~{\bf 10}, 282–285 (2011).

\bibitem{for13}
F. Forster, A. Molina-Sanchez, S. Engels, A. Epping, K. Watanabe, T. Taniguchi, L. Wirtz, and C. Stampfer, Phys. Rev. B,~{\bf 88}, 085419 (2013)

\bibitem{bis12}
D. Bischoff, T. Kr\"ahenmann, S. Dr\"oscher, M. A. Gruner, C. Barraud, T. Ihn, and K. Ensslin
Appl. Phys. Lett.,~{\bf 101}, 203103 (2012).

\bibitem{eng13}
S. Engels, A. Epping, C. Volk, S. Korte, B. Voigtl\"ander, K. Watanabe, T. Taniguchi, S. Trellenkamp, and C. Stampfer
Appl. Phys. Lett.,~{\bf 1003}, 0731133 (2013).



\bibitem{tom11}
N. Tombros, A. Veligura, J. Junesch, M. H. D. Guimaraes, I. J. Vera-Marun, H. T. Jonkman, and B. J. van Wees, Nat. Phys.,~{\bf 7}, 697 (2011)

\bibitem{mor12}
Dong-Keun Ki and A. F. Morpurgo, Phys. Rev. Lett.,~{\bf 108}, 266601 (2012). 


\bibitem{wan11}
X. Wang, Y. Ouyang, L. Jiao, H. Wang, L. Xie,	J. Wu, J. Guo	and H. Dai, Nat. Nanotechnol.,~{\bf 6}, 563 (2011).



\bibitem{dav07a} 
A. C. Ferrari, J. C. Meyer, V. Scardaci, C. Casiraghi, M. Lazzeri, F. Mauri, S. Piscanec, D. Jiang, K. S. Novoselov, S. Roth and A. K. Geim, Phys. Rev. Lett.,~{\bf 97}, 187401 (2006). 
D. Graf, F. Molitor, K. Ensslin, C. Stampfer, A. Jungen, C. Hierold, and
L. Wirtz, Nano Lett., {\bf 7}, 238 (2007).


\bibitem{com02}
Here we follow Molitor et al. \cite{mol10}.
For extracting the transport gap the data is smoothed with a running average over 0.5~V in $V_{\rm bg}$ and the regions with a linear increase in conductance next to the transport gap are fitted with a linear slope. From the intersection of these two (right and left) lines with $G=0$ we obtain $V_r$ and  $V_l$ and extract the size of the transport gap $\Delta V_{\rm bg}=(V_r-V_l)$ and the position of the charge neutrality point $V_{\rm cn}=(V_r+V_l)/2$ . For the effective energy gap the size of the largest Coulomb diamond like feature in a plot of $dI/dV$ as a function of bias and back gate voltage is estimated.

\bibitem{footnote1}
This also confirms that our GNRs are not suspended; when fully suspending 
individual nanoribbons we found $\alpha_a \approx 0.05$. 

\bibitem{sol07}
F.~Sols, F.~Guinea and A.~H.~Castro-Neto, Phys. Rev. Lett.,~{\bf 99}, 166803 (2007).

\bibitem{comCS01}
In Ref.~\cite{mol10} the following values for $a$ and $b$ have been found: $a=2.0$~eV~nm and $b=0.026$~nm$^{-1}$.

\bibitem{du08}
X. Du, I. Skachko, A. Barker, and E.~Y. Andrei, Nat. Nanotechnol.,~{\bf 3}, 491 (2008).

\bibitem{bol08}
K. I. Bolotin, K. J. Sikes, Z. Jiang, M. Klima, G. Fudenberg, J. Hone, P. Kim, H. L. Stormer, Solid State Commun.,~{\bf 146}, 351 (2008).


\bibitem{comm03}
The estimated values are $C_0/C_{\rm bg}=24$ and $\gamma=8.5 \times 10^{-4}$~V$^{-2}$ ($C^{(1)}_{\rm sg}/C_{\rm bg}=1.15$).

\bibitem{tac10} H. Tachikawa, T. Iyama, and H. Kawabata,
Japanese Journal of Applied Physics 49 (2010) 01AH02.

\bibitem{ihna11}
S. Ihnatsenka, and G. Kirczenow, Phys. Rev. B,~{\bf 83}, 245442 (2011).
 
\bibitem{zhe08}
H. Zheng, and W. Duley, Phys. Rev. B,~{\bf 78}, 045421 (2008).

 
\end{thebibliography}
\end{document}